# Twisted graphene bilayer around the first magic angle engineered by heterostrain

Jia-Bin Qiao[1], Long-Jing Yin[1,2] and Lin He[1,*]

[1]Center for Advanced Quantum Studies, Department of Physics, Beijing Normal University, Beijing 100875, People's Republic of China
[2]School of Physics and Electronics, Hunan University, Changsha 410082, People's Republic of China
[*]Corresponding author. (E-mail: helin@bnu.edu.cn).

**Very recently, twisted graphene bilayer (TGB) around the first magic angle $\theta \approx 1.1°$ has attracted much attention for the realization of exotic quantum states, such as correlated insulator behavior and unconventional superconductivity. Here we elaborately studied a series of TGBs around the first magic angle engineered by heterostrain, where each layer is strained independently. Our experiment indicated that a moderate heterostrain enables the structural evolution from the small-angle TGB ($\theta \sim 1.5°$) to the strained magic-angle TGB ($\theta \sim 1.1°$), exhibiting the characteristic low-energy flat bands. The heterostrain can even drive the system into highly strained tiny-angle TGBs ($\theta \ll 1.1°$) with large deformed tetragonal superlattices, where a unique network of topological helical edge states emerges. Furthermore, the predicted domain wall modes, which are strongly localized and result in hexagon-triangle-mixed frustrated lattice derived from Kagome lattice, are observed in the strained tiny-angle TGBs.**

Strain in monolayer two-dimensional (2D) materials introduces spatially varying lattice deformations and consequently modulates intralayer hopping strength, giving rise to rich emergent phenomena in these systems [1-9]. In bilayer 2D materials with weak interlayer van der Waals bonding, for example graphene bilayer, strain may lead to even more exotic phenomena because lattice deformation can further change the atomic alignment between the adjacent layers, which strongly affects interlayer electron motion [10-14]. This is especially the case in twisted graphene bilayer (TGB) because that the electronic properties are sensitively depending on the twisted angle $\theta$ between the layers [15-33]. Recently, Mott insulating behavior [27] and unconventional superconductivity [28] have been observed in the TBG at the first magic angle $\theta \approx 1.1°$, which attract enormous attention. In this Letter, we systematically studied effects of heterostrain on a series of TGB around the first magic angle [16-18,27-31] *via* scanning tunneling microscopy/spectroscopy (STM/STS) measurements, and demonstrated that an in-plane multiaxial heterostrain could dramatically change both geometric structures and electronic properties of the system. The variations of the atomic alignment between the layers induced by the heterostrain control the physics of the system and allow access to previously inaccessible structures and exotic quantum states in the TGB.

Figure 1(a) shows a schematic image depicting effects of a representative heterostrain on the stacking order of the graphene bilayer. The initial structure of the bilayer is Bernal-stacked. For simplicity, the topmost layer is a pristine graphene sheet and there is a non-uniform biaxial in-plane strain $\varepsilon$ along both the zigzag and armchair directions in the underlying graphene sheet. Obviously, the heterostrain changes the stacking order from the AB-stacked mode and introduces a relative twist, which varies gradually from the left to the right, between the adjacent layers, as shown in Fig. 1(a). In the presence of a twist, the magnifying effect of the moiré pattern [34-36] enables us to clearly image and directly measure a small heterostrain in graphene bilayer. In our experiment, the studied graphene bilayer is on top of α-$Mo_2C$ and the bilayer graphene/α-$Mo_2C$ heterostructures is synthesized on Cu foils *via* a one-step chemical vapor deposition (CVD) method (see Supplemental Materials Methods, fig. S1 and note S1 for more details [37]). The growth of bilayer graphene on α-$Mo_2C$ is driven by the high carbon precursor supersaturation, which is analogous to that of few-layer graphene synthesized on copper foils [38]. Owing to the strong interaction and lattice mismatch between graphene and $Mo_2C$ [39,40], the underlying graphene sheet is easily subjected to the strain generated by the supporting α-$Mo_2C$, while the topmost sheet remains pristine graphene, as schematically shown in Fig. 1(b).

Figure 1(c) shows a STM image of a representative strained graphene bilayer area observed in our experiment, exhibiting extreme changes of the moiré patterns. According to the measured structure, we can roughly divide the studied area into three regions. In region I, we observed almost unstrained triangular moiré patterns with the period $D \approx 10.7$ nm. Figure 1(d) shows six well-defined hexangular spots as the moiré reciprocal lattice. In region II, the period of the moiré patterns increases and becomes anisotropy. Therefore, we observed deformed triangular moiré patterns and deformed hexangular spots in the reciprocal lattice shown in Fig. 1(e). In contrast to the triangular superlattices observed in regions I and II, the moiré patterns in region III turn into deformed tetragonal lattices with much larger moiré wavelengths. The deformed tetragonal lattices are further verified by the moiré reciprocal lattices shown in Fig. 1(f).

According to our atomic-resolved STM measurement (see Supplemental Materials fig. S2 [37]), there is undetectable lattice deformation of the topmost graphene sheet in the structure shown in Fig. 1(c). The observed extreme changes of the moiré patterns are hence attributed to the in-plane lattice deformation of the underlying graphene sheet generated by the supporting substrate. To further confirm the above analysis, we carried out the sketchy simulation on effects of the heterostrain on the structures of moiré patterns. In the TGB, a complex multiaxial heterostrain can be approximately equivalent to the superposition of a uniaxial strain and a relative twist (see Fig. S3 and S4 and note S2 [37]). Therefore, in our simulation, we assume that the topmost graphene layer of the TGB is intact, while a non-uniform in-plane uniaxial strain along the armchair direction of the second graphene sheet and there is a relative twist between the adjacent sheets. Figure 1(g) shows the simulated result with both the gradually increasing uniaxial strain and decreasing twist angle from the left to right (the strain-induced relative twist is opposite to the initial twist angle). Obviously, our simulation grasps the main morphological features observed in our experiment, i.e., it reproduces the three different regions of the moiré patterns. Here it should be pointed out that the large-scale strained moiré patterns are not observed by chance but, however, always found out on the whole heterostructure surface and featured by continuously and gradually varied moiré periods (see Fig. S5 [37]), unveiling the ubiquity of the heterostrain in the system.

In the TGB, the moiré pattern could magnify small lattice deformations of graphene sheets and a very small heterostrain will result in large variations of the moiré patterns [34-36]. In the deformed triangular moiré regions, the in-plane heterostrain could lead to a striking structural transition from the almost unstrained ($\varepsilon \sim 0\%$) small-angle TGB with the average period about 9.5 nm ($\sim 1.5°$) to the strained ($\varepsilon \sim 0.3\%$) "magic-angle" one with the average period around 13.5 nm ($\sim 1.1°$) (see Fig. S6 and note S2 [37]). Moreover, there exist negligible electronic coupling between the TGB and $Mo_2C$ substrate underneath (see Fig. S7 [37] for STM/STS results of monolayer graphene on $Mo_2C$). Therefore, our system provides an excellent platform to investigate the effect of heterostrain on the geometric structures and electronic properties of the TGB around the first magic angle. Figure 2(a) shows a representative STM image of the TGB region with the average period of the moiré patterns changing from about 8.86 nm to about 12.45 nm. The corresponding twist angle decreases from about 1.45° to about 1.13°, which is quite close to the first magic angle [16-18,27-31]. It is easy to identify a transition area of two deformed triangular moiré unit cells with different periods. Our STS measurements, i.e., d$I$/d$V$ spectra, observed two low-energy van Hove singularities (VHSs) in the 1.43° moiré pattern (see Supplemental Materials fig. S8 [37]), stemming from the emergence of two saddle points at M point in the electronic structure as shown in left panel of Fig. 2(b) [19-24]. The energy separation of the two VHSs agrees with that reported previously in the unstrained TGB with a similar twist angle [30]. In the first "magic-angle" ($\sim 1.13°$) moiré pattern, the two low-energy VHSs merge into a single pronounced peak (see Supplemental Materials fig. S8 [37]), which is the characteristic feature of the flat bands (FB) in the magic-angle TGB, as shown in right panel of Fig. 2(b) [30]. To clearly show the evolution of the electronic properties of the deformed moiré patterns, we plotted $d^2I/dV^2$ as a function of measured positions in Fig. 2(c). Obviously, the two VHSs abruptly merge into the single pronounced peak of the FB when we moved the STM tip to the first magic-angle moiré pattern.

To directly visualize the spatial variations of electronic properties in the deformed TGB, we carried out differential conductance maps (STS maps), which reflect the spatial distribution of the local density-of-states (LDOS) at the measured energy. There are three primary features in the LDOS maps. Firstly, at the energies where the FB and the VHSs reside, the signals of LDOS are much stronger in the AA-stacked regions of the moiré pattern, as shown in Fig. 2(d) and (e) respectively, indicating that the Dirac fermions are highly confined in the AA-stacked regions. We can clearly distinguish the magic-angle TGB region from the other TGB region with different twist angles owing that the LDOS is much stronger in the magic-angle TGB region at the energy of the FB (Fig. 2(d)). Secondly, even though the LDOS signals are stronger in the AA regions at both the VHSs and the FB peaks, the signals become stronger in the AB and BA regions at high energy, as shown in Fig. 2(f). Such a result may be related to the energy-dependent wave function distribution of the moiré bands in the TGB [31,41,42]. Finally, we observed the inhomogeneous LDOS distribution in the non-magic-angle regions where the twist angles are almost the same (~ 1.4 °) (see Fig. 2(d), 2(e), Supplemental Materials fig. S9 and note S3 [37]), unveiling the remarkable modulation of heterostrain on the electronic properties of the TGB.

A relatively large heterostrain can further drive the triangular moiré pattern of the TGB into the deformed tetragonal moiré superlattices with periods much larger than the period of the first magic-angle TGB, corresponding to the tiny angles ($\theta \ll 1.1°$). Figure 3(a) shows a representative STM image of the tetragonal moiré superlattices. The bright spots in topography are the AA-stacked regions and the distance between two adjacent moiré pattern ranges from about 18.2 nm to about 35.2 nm, corresponding to the twist angles from ~ 0.78° to ~ 0.40°. Inside the moiré unit cells, two distinct stacking types are directly visualized in the atomic-resolution STM topography, as shown in Fig. 3(b) and (c). One exhibits a clear triangular lattice (Fig. 3(b)), indicating the AB/BA-stacked mode. The other, which connects two adjacent AB and BA regions, displays the AA-stacked hexagonal lattice (Fig. 3(c)). Such a result reminds the AB-BA domain wall (DW) observed recently in Bernal graphene bilayer [11-13] and very recently in the TGB with twist angle much smaller than the first magic-angle [32,33].

Our STS measurements in the deformed tetragonal moiré pattern indicate that the electronic properties depend sensitively on the stacking types within the moiré unit cell, as shown in Fig. 3(d) (see also Fig. S10 [37]). The spectra recorded on the AA-stacked regions show a pronounced resonance at Dirac point $E_D$ (~ - 21 meV). It is reasonable to attribute it to the AA modes arising from the particular moiré bands confined in the AA regions [43]. Note that the Dirac energy $E_D$ in the deformed tetragonal moiré pattern is slightly lower than that in the adjacent "magic-angle" TBG systems ($E_D$ ~ - 10 meV), which may arise from extra electron doping induced by the enhanced next-nearest-neighbor hopping in strained graphene [44]. The tunneling spectra taken on the AB/BA-stacked regions are characterized by a finite gap ($E_g \sim 230$ meV) induced by the (substrate-induced) interlayer electric field, while there emerge in-gap states in the spectra obtained in the DW regions. Such a result recalls the scenario occurring around the AB-BA domain walls where an inversion-symmetry-broken gap emerges in both the AB and BA regions and the topological helical edge states appear in the domain walls and form the helical network [12,13,32,33,45,46]. Moreover, there are several resonances in the spectra acquired in the DW regions, and the intensity of the resonances becomes weaker in the AB/BA regions and almost vanishes in the AA regions. Similar phenomena have also been observed in many deformed tetragonal moiré regions with different substrate-induced electric fields (see Fig. S11 [37]). We attribute the peaks in the spectra to the spatially localized domain wall (DW) modes, or pseudo Landau levels (pLLs), arising from the interlayer electric bias induced gauge field [47]. Figure 3(e) shows the energy of the peaks as a function of the distinct gaps (i.e., different electric fields) observed in our experiment, which is consistent with the theoretically predicted evolution of the DW modes depending on the applied electric fields (see Fig. 2 and 4 in Ref. [47]). All the results indicate the coexistence of the helical network and DW modes in the tiny-angle TGB.

To further confirm the above experimental results, we directly imaged these nontrivial states by operating energy-fixed STS maps. Figures 4(a,b) show STS maps recorded at different energies. For STS maps recorded within the gap, the electronic states at the AA-stacked regions are connected by quasi-one-dimensional conducting channels to form an unusual network exhibiting mutually splicing triangle- and inverted triangle-shaped configuration (Fig. 4(a)). In STS maps taken at the energy of the DW modes, surprisingly, the LDOS are quite localized in the DW and form an unusual hexagon-triangle-mixed lattice surrounding the AA regions (Fig. 4(b)).

To better understand effects of the heterostrain on the helical network and the DW modes, we operated cartoon simulations on geometric structures and LDOS distributions of the tiny-angle TBG with interlayer electric bias. In the presence of electric fields, topological edge states reside on the DW regions and form the network, as shown in Fig. 4(c) and (e). Intriguingly, the strain-modulated change of network configuration could be treated as a motion of the network unit cell (labelled by yellow dashed triangle) in unstrained supercells following the strain-induced shift of AA regions. When it turns to the DW modes, the helical channels on the DW regions in unstrained superlattice degrade into highly localized states and then form a Kagome lattice, as shown in Fig. 4(d), which may be closely linked to quantum spin liquids and unconventional superconductivity [48-50]. Similarly, the DW modes are substituted for the unique network in strained superlattice and constitute a new hexagon-triangle-mixed frustrated lattice derived from the Kagome lattice shown in Fig. 4(f).

In summary, we systematically studied geometric structures and electronic properties of the heterostrain-engineered TGB around the first magic angle. We demonstrated that a slight heterostrain dramatically enables twist tunability and electronic structure modulation in graphene bilayer system. Our findings not only pave a new way to realize previously inaccessible twisted van der Waals heterostructures based on heterostrain engineering, but also provide an unprecedented platform to study the interplay between different nontrivial states, involving topological states and many-body states.

**Acknowledgements**

This work was supported by the National Natural Science Foundation of China (Grant Nos. 11674029, 11422430, 11374035), the National Basic Research Program of China (Grants Nos. 2014CB920903, 2013CBA01603). L.H. also acknowledges support from the National Program for Support of Top-notch Young Professionals, support from "the Fundamental Research Funds for the Central Universities", and support from "Chang Jiang Scholars Program".

**Figures**

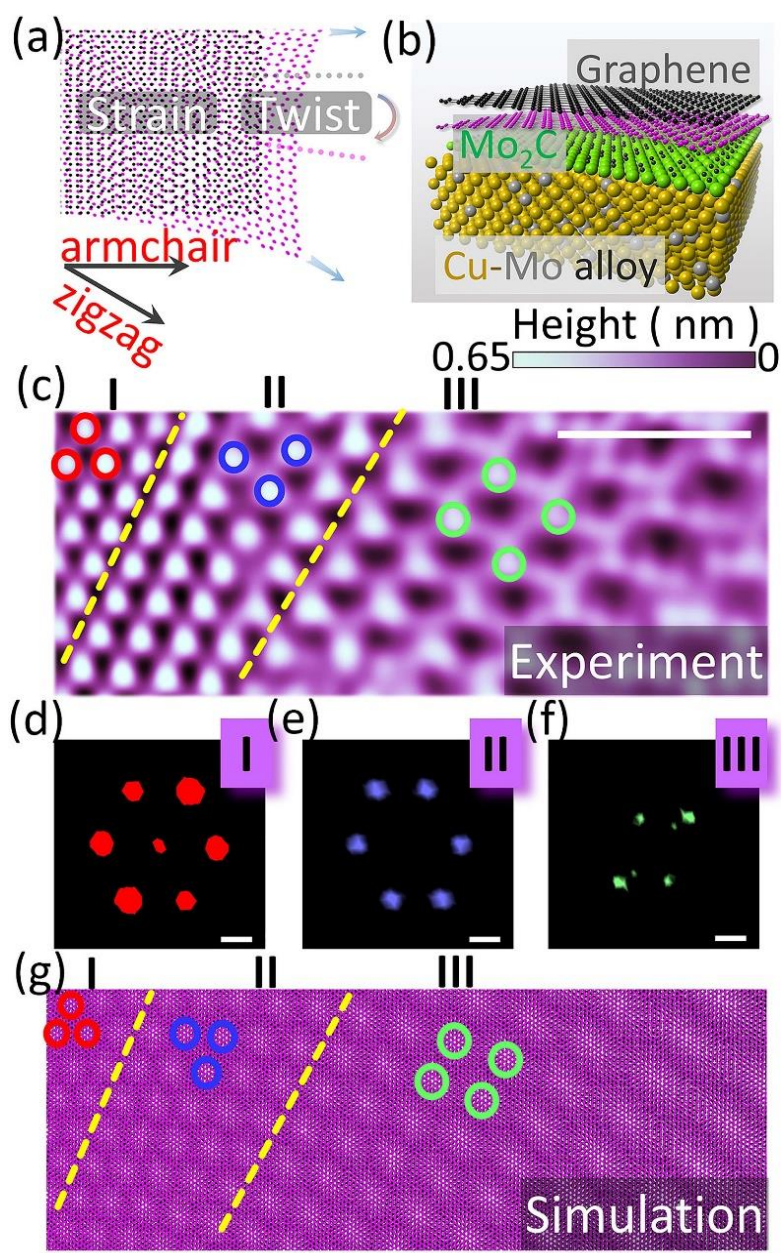

**Figure 1** (**a**) Cartoon for the structural transition of bilayer graphene from AB-stacked mode to rotation-like stacking faults induced by a non-uniform biaxial strain along the zigzag and armchair directions. (**b**) Schematic of one-step-grown bilayer graphene/α-Mo$_2$C heterostructures. (**c**) STM topography ($V_S$ = 300 mV, $I$ = 0.2 nA) of TGB with different moiré superlattices induced by the heterostrain. Topo I: almost unstrained triangular moiré superlattice (marked by red circles); II: deformed triangular moiré

superlattice (blue circles); III: deformed tetragonal moiré superlattice (green circles). Scale bars, 50 nm. (**d-f**) FFT images of three typical moiré superlattices in the Topo I, II and III, respectively. Scale bars, 0.05 nm$^{-1}$. (**g**) Simulation of the structural transition in panel (**c**) based on the reduced model where the underlying graphene sheet encounters a gradually increasing (non-uniform) uniaxial strain along the armchair direction and a progressively decreasing twist angle relative to the unstrained topmost layer.

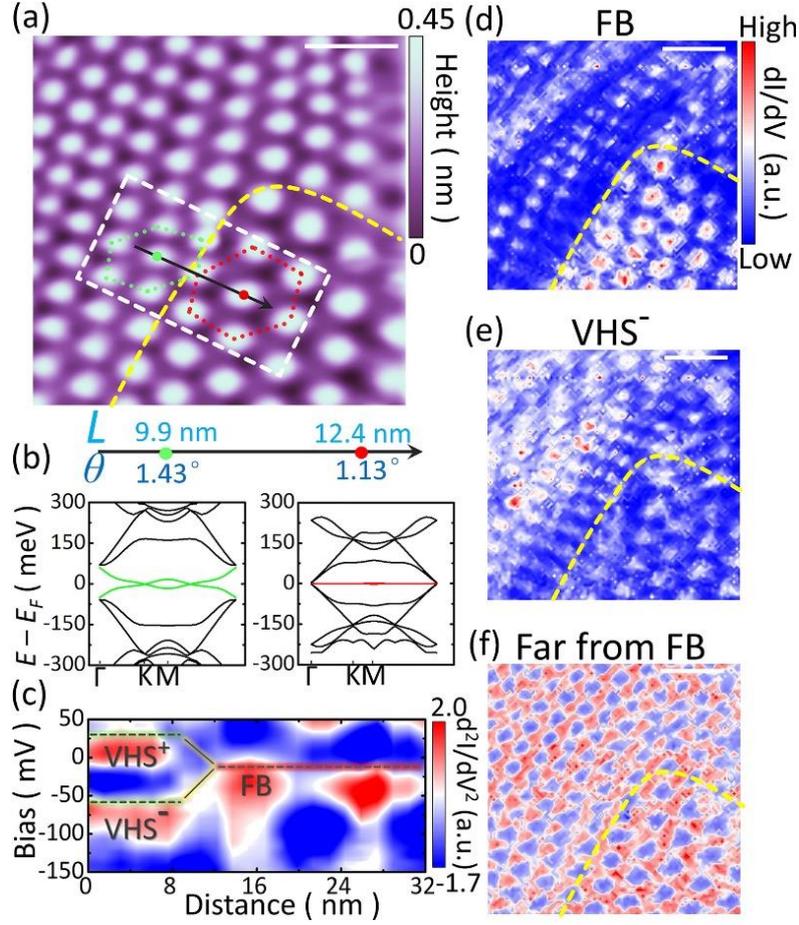

**Figure 2** (**a**) STM morphology ($V_S$ = 200 mV, $I$ = 0.2 nA) of deformed triangular moiré superlattices with effective twist angles near the first "magic angle". Areas for magic-angle TGB and the other TGB are roughly divided by a yellow dashed curve. Scale bar, 20 nm. A transition area is labelled by the white dashed frame. The two moiré patterns with different periods are marked by deformed green and red dotted hexagons, respectively. (**b**) Theoretical calculated electronic structures of the TBG with twist angle $\theta$ ~ 1.43 ° and ~ 1.1 °, respectively. (**c**) $d^2I/dV^2$ map measured along the black arrow in panel (**a**). The positions of the VHSs (VHS$^+$ and VHS$^-$) and the FB are marked with black dashed lines. (**d-f**) $dI/dV$ maps obtained over the whole areas in panel (**a**) at the energy of the FB (-9 meV), VHS$^-$ (-50 meV), and away from the FB (- 300 meV), respectively. Scale bars, 20 nm.

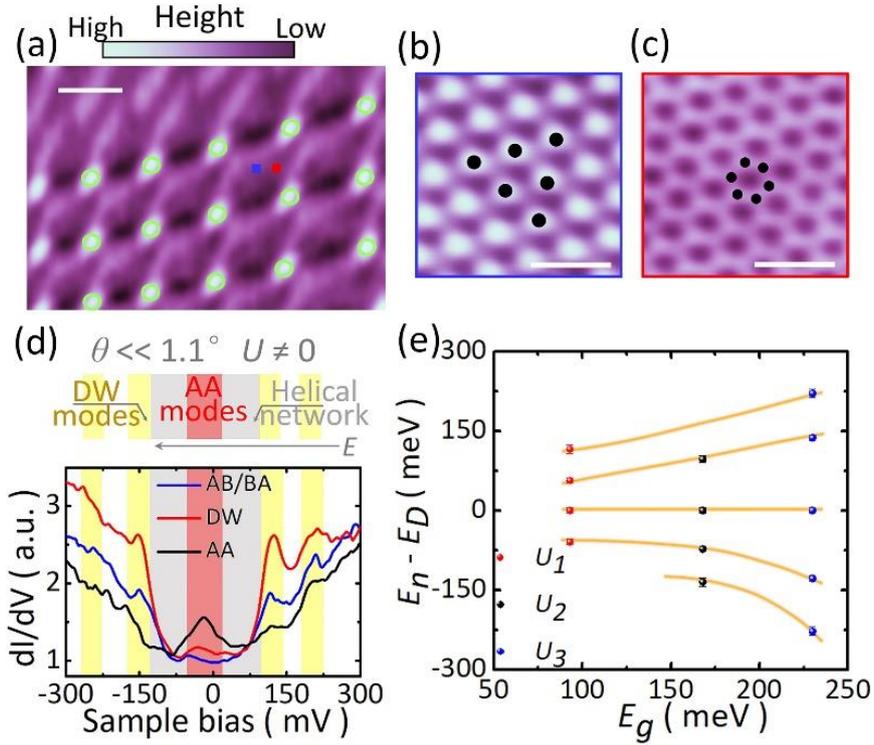

**Figure 3** (**a**) STM topography ($V_S$ = 200 mV, $I$ = 0.2 nA) of deformed tetragonal moiré supercells. Scale bar, 20 nm. (**b** and **c**) Atomic-resolution STM images ($V_S$ = 700 mV, $I$ = 0.2 nA) of the AB/BA-stacked and domain wall regions labelled by the blue and red dots in panel (**a**), displaying triangular and hexagonal lattices, respectively. Scale bars, 0.5 nm. (**d**) Upper panel: Schematic of the spectra of the tiny-angle TGB under non-zero external electrical bias $U$; Lower panel: Representative STS spectra taken at the AB/BA, AA and domain wall regions. The FB resonance, helical network and DW modes is marked. (**e**) The energy of pLL-like DW modes $E_n − E_D$ deduced from the spectra acquired on three different tetragonal moiré regions under different electric fields $U_i$ ($i$ = 1,2,3) as a function of the corresponding gaps $E_g$ induced by the fields.

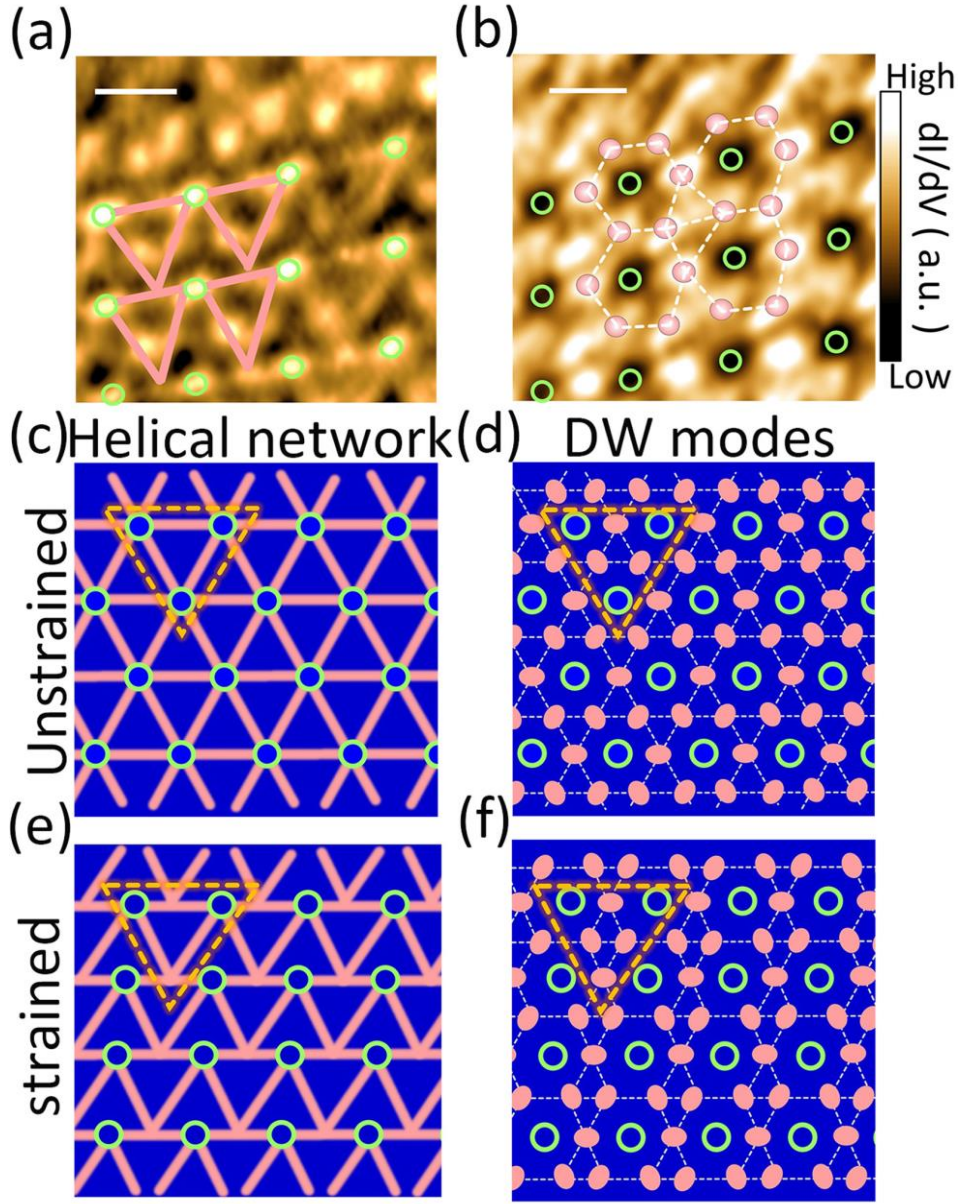

**Figure 4** (**a,b**) *dI/dV* maps obtained over the whole regions in Fig. 3(a) with the fixed sample bias of -46 mV (**a**), and 200 mV (**b**), respectively. Scale bars, 20 nm. Topological networks (denoted by rose lines) are formed on the DW regions in the map within the gap, while an exotic hexagon-triangle-mixed lattice formed by the DW modes is observed in the map obtained at a resonant level. (**c-f**) Cartoons of unstrained and strained networks (**c** and **e**, marked by rose lines) and DW modes (**d** and **f**, marked by rose spots), respectively.